\documentclass[runningheads]{llncs}
\usepackage[T1]{fontenc}
\usepackage{graphicx}
\usepackage{xcolor}
\usepackage[normalem]{ulem}
\newcounter{commentCounter}

\newcounter{RZNumberOfComments}
\stepcounter{RZNumberOfComments}

\begin{document}
\title{Are We There Yet? Unravelling Usability Challenges and Opportunities in Collaborative Immersive Analytics for Domain Experts}
\titlerunning{CIA: Usability Challenges and Opportunities}
%
\author{Fahim Arsad Nafis\inst{1}\orcidID{0009-0005-5803-5200} \and
Alexander Rose \inst{1}\orcidID{0009-0004-1915-4357} \and
Simon Su \inst{2}\orcidID{0000-0002-2460-3899} \and
Songqing Chen \inst{1}\orcidID{0000-0003-4650-7125} \and
Bo Han \inst{1}\orcidID{0000-0001-7042-3322}}
\authorrunning{F. Nafis et al.}
%
\institute{George Mason University, Fairfax VA 22030, USA \email{\{fnafis2, arose23, sqchen, bohan\}@gmu.edu} \url{https://www.gmu.edu/} \and
Montgomery College, Rockville MD 20850, USA \email{simon.su@montgomerycollege.edu} \url{https://www.montgomerycollege.edu/} }
\maketitle              
\begin{abstract}
In the ever-evolving discipline of high-dimensional scientific data, collaborative immersive analytics (CIA) offers a promising frontier for domain experts in complex data visualization and interpretation. 
This research presents a comprehensive framework for conducting usability studies on the extended reality (XR) interface of ParaView, an open-source CIA system.
By employing established human-computer interaction (HCI) principles, including \textit{Jakob Nielsen's Usability Heuristics}, \textit{Cognitive Load Theory}, \textit{NASA Task Load Index}, \textit{System Usability Scale}, \textit{Affordance Theory}, and \textit{Gulf of Execution and Evaluation}, this study aims to identify underlying usability issues and provide guidelines for enhancing user experience in scientific domains.
Our findings reveal significant usability challenges in the ParaView XR interface that impede effective teamwork and collaboration.
For instance, the lack of synchronous collaboration, limited communication methods, and the absence of role-based data access are critical areas that need attention. 
%
Additionally, inadequate error handling, insufficient feedback mechanisms, and limited support resources during application use require extensive improvement to fully utilize the system’s potential.
%
%
Our study suggests potential improvements to overcome the existing usability barriers of the collaborative immersive system.


\end{abstract}
\section{Introduction}
\begin{figure}
\centering
\includegraphics[scale=0.35]{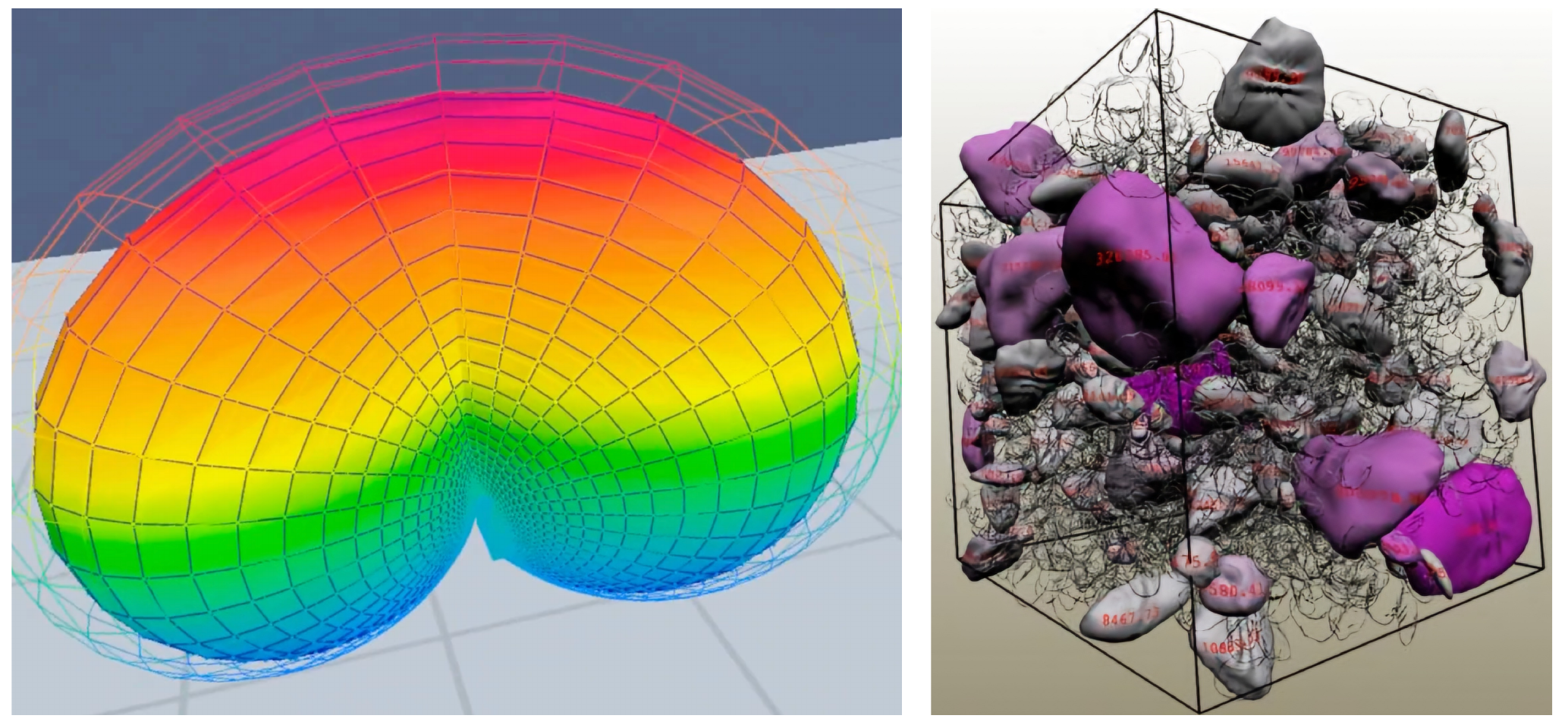}
\caption{Visualization of 3D scalar active region magnetic field (left), visualization of the concrete modelling and simulation on a high-performance computer (right).}
\label{intro_1}
\end{figure}
Visualizing high-volume scientific data is crucial in a wide range of domains, such as space weather
forecasting~\cite{angryk2020multivariate,zhang_earthaffecting_2021}, medical imaging~\cite{alzubaidi2021novel,shamshad2023transformers}, and 
high-performance computing (HPC)~\cite{isaacs2014state,del2020gauge}, as shown in Figure~\ref{intro_1}.
Immersive analytics (IA)~\cite{marriott2018immersive} is a groundbreaking way to drastically improve engagement with complex datasets by granting six degrees of freedom (6DoF) movement for the users~\cite{kraus2022immersive,IA_6DOF_2022}. 
As the next generation of immersive analytics, collaborative immersive analytics (CIA)~\cite{billinghurst2018collaborative} has emerged as a transformative approach to data exploration and decision-making. Compared to the traditional IA approach, the CIA empowers domain experts to dive deeper into complex datasets, allowing them to gain richer insights and facilitating interdisciplinary collaboration.

\begin{figure}
\centering
\includegraphics[width=\textwidth]{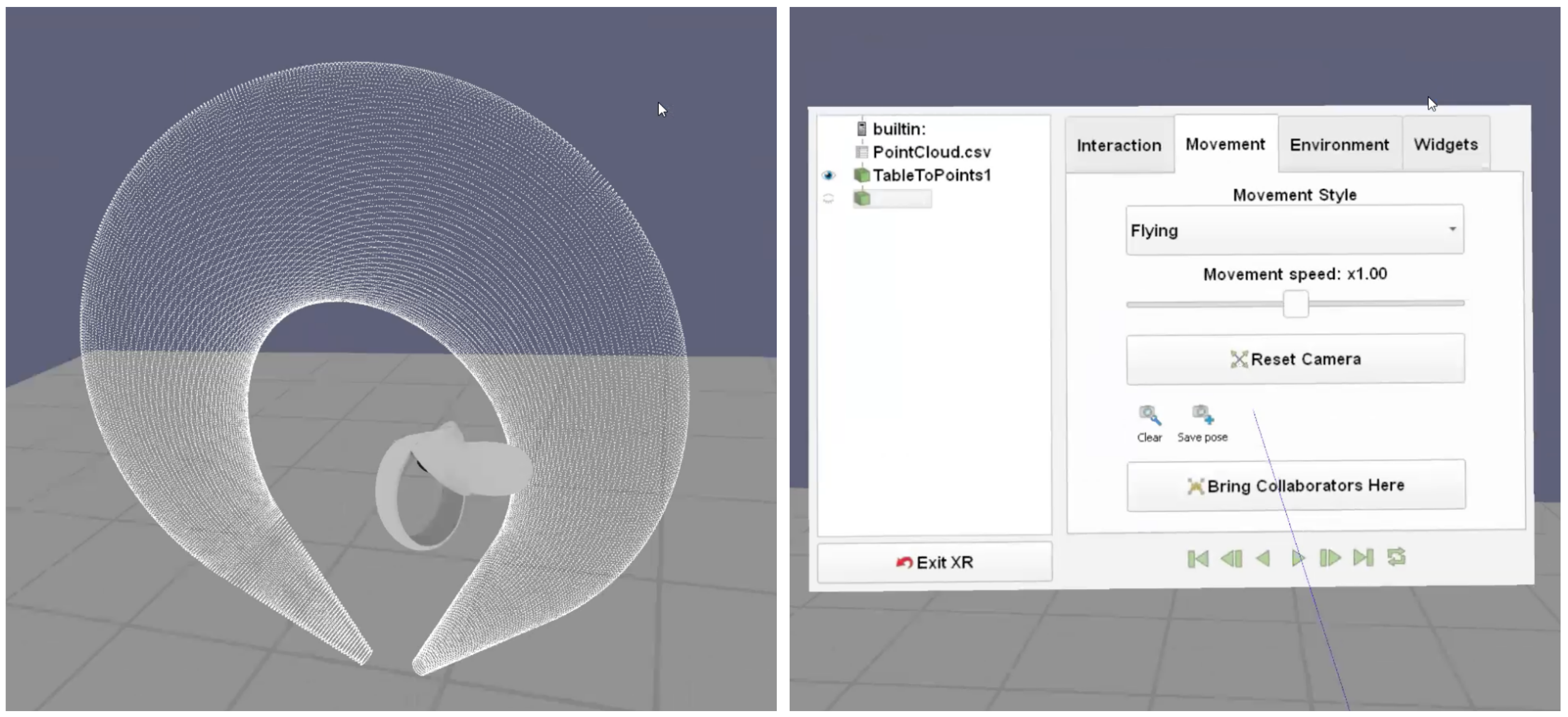}
\caption{Interaction with the visualization of 3D vector dipole magnetic field in ParaView XR interface (left), a screenshot of ParaView XR interface user menu (right).} 
\label{intro_2}
\end{figure}

In this paper, we investigate ParaView~\cite{ahrens200536,paraview}, a popular platform to visualize scientific data.
%
%
Specifically, we examine the ParaView XR interface, as shown in Figure~\ref{intro_2}, to explore its challenges and opportunities in usability in the context of CIA.
%
%
It extends the traditional capabilities of ParaView by incorporating virtual reality (VR)~\cite{burdea2003virtual} and augmented reality (AR)~\cite{azuma1997survey} technology, allowing users to freely navigate and interact with large-scale scientific datasets~\cite{su2022immersive}. 
Its significance lies in facilitating collaborative exploration and decision-making among multiple domain experts.
%
%

The potential of the CIA to revolutionize the scientific workflow is immense, yet several usability challenges hinder its full integration~\cite{chen2021effect,ens2021grand,reski2022empirical}. Different from traditional interfaces, immersive environments require users to engage with the data through diverse methods. It poses unique challenges in collaborative settings where multiple users must effectively share and manipulate data, yet potentially without physical cues.
%
%
Although ParaView has significantly advanced the integration of XR features in data visualization, the usability of these tools in collaborative scientific environments remains uncharted~\cite{isenberg2011collaborative,kerren2012toward}. Therefore, conducting a usability study in this virtual environment is crucial to identifying and overcoming barriers to effective collaboration and data exploration. 

Specifically, our study aims to answer the following four research questions (RQs), designed to meticulously examine multiple aspects of usability and collaborative efficacy of the ParaView XR interface:

\begin{itemize}
    \item \textbf{RQ1: }How easily can domain scientists individually learn and utilize the immersive visualization capabilities of ParaView?
    \item \textbf{RQ2: }How effective are the collaborative tools within ParaView when used in a multi-user immersive environment by domain scientists?
    \item \textbf{RQ3: } What usability challenges do domain scientists encounter while using the ParaView XR interface in a collaborative setting?
    \item \textbf{RQ4: } What improvements or additional features can be implemented to enhance the collaborative user experience of the ParaView XR interface?
\end{itemize}

In this research, we present a comprehensive framework for conducting usability studies on CIA systems. 
Specifically, we make the following contributions:
%
\begin{itemize}
    \item Design an experimental framework tailored specifically for usability studies in CIA by utilizing established principles in the HCI area.
    \item Conduct a pilot study to validate the effectiveness of this framework.
    \item Investigate the usability issues of the ParaView XR interface and analyze its user experience in individual and collaborative contexts.
    \item Provide recommendations and enhancements for CIA systems to improve their usability and user experience.
\end{itemize}

\section{Background}
\subsection{Collaborative Immersive Analytics}
Immersive analytics is an emerging field that combines data visualization and analytics with immersive technologies, such as VR and AR~\cite{marriott2018immersive}. Utilizing human natural spatial skills by presenting data in three-dimensional settings makes exploring and comprehending complex datasets easier. This approach promotes a more intuitive understanding by visually representing information in an engaging immersive environment~\cite{bach2016immersive,fonnet2019survey}. Several research directions in IA are actively being pursued to enhance usability. Recent studies systematically review the design space of IA~\cite{saffo2023unraveling}, introduce a toolkit to facilitate the development of IA applications~\cite{cordeil2019iatk}, and provide a web-based framework to simplify the creation of IA experiences~\cite{butcher2020vria}.

CIA extends the paradigm of IA by allowing multiple users to simultaneously visualize, and interact with data within shared virtual environments. 
A recent study shows that the use of shared surfaces and spaces can enhance collaborative data visualization in immersive co-location environments~\cite{lee2020shared}. Another investigation focuses on the design of collaborative frameworks, aiming to improve teamwork and data interaction within the virtual environment~\cite{nguyen2016collaborative}. 
A recent pilot study examines how different modes of collaboration and positional arrangements affect user performance on IA tasks in VR~\cite{chen2021effect}. 

\subsection{Human-computer Interaction Principles}
In this section, we introduce several established principles of HCI that are fundamental to designing and evaluating user interfaces for this study.

\vspace{0.05in}
\noindent {\bf Jakob Nielsen’s 10 Heuristics~\cite{nielsen1994usability}}
are a widely adopted set of principles for assessing and improving user interface design. This principle covers multiple aspects of a system to examine its efficacy. These heuristics cover aspects, such as visibility of system status, the match between the system and the real world, user control and freedom, consistency and standards, error prevention, recognition rather than recall, flexibility and efficiency of use, aesthetic and minimalist design, help for users to recognize and recover from errors, and adequate help and documentation. 
%
%
Jakob Nielsen's heuristic principle has been extensively applied in various studies~\cite{deyoung2018evaluating,wang2019usability} assessing the usability of virtual environments, making it a suitable choice for evaluating the experiments in this study.

\vspace{0.05in}
\noindent {\bf Cognitive Load Theory (CLT)~\cite{sweller1998cognitive}} explains how information processing demands can affect a user's ability to perform tasks effectively. According to this theory, human memory can be divided into three main types of cognitive load: intrinsic, extraneous, and germane~\cite{plass2010cognitive}. The intrinsic load relates to the complexity of the learning material itself. The extraneous load comes from how the information is shown, which can make learning harder if it is not done well. The german load involves mental activities that help to understand and organize new information. This theory can help in understanding how immersive environments impact the cognitive processing of complex datasets. Several studies~\cite{haryana2022virtual,souchet2022measuring} use this theoretical strategy to minimize unnecessary cognitive load improving learning and data comprehension.


\vspace{0.05in}
\noindent {\bf Affordance Theory~\cite{gibson1977theory}} describes how an object's perceived properties influence its usability. It is relevant in VR and AR environments where user interaction modes are not standardized. When a system includes perceptible affordances, users find its features easy to use. Hidden affordances keep users unaware of certain features resulting in a less effective user experience. Users face false affordances of the system while interacting with elements that misleadingly suggest functionality.

\vspace{0.05in}
\noindent {\bf System Usability Scale (SUS)~\cite{bangor2008empirical}} provides a reliable tool to measure the usability of a system. It offers quantitative data that help to assess how users can utilize these advantages in practice. Numerous studies~\cite{chandra2019review,wijaya2019usability} have used SUS to assess the usability of VR platforms in formal research.

\vspace{0.05in}
\noindent {\bf Gulf of Execution and Evaluation~\cite{norman1986cognitive}} addresses the gap between users' intentions of a system and their actual experiences. 
The gulf of execution refers to the challenge of determining the necessary steps to achieve desired outcomes. The gulf of evaluation involves understanding the current state of the system and how it aligns with the user's objectives. Recent studies utilize this concept to analyze usability across diverse applications, such as 
VR video editing~\cite{nguyen2017vremiere}, and mediating human-robot interactions through mixed reality~\cite{szafir2019mediating}.

\vspace{0.05in}
\noindent {\bf NASA Task Load Index (TLX)~\cite{hart1988development}} is a widely used tool that measures workload to assess the user experience. It evaluates factors, such as mental, physical, and temporal demand, satisfaction, effort and frustration caused by performing a certain task~\cite{hart2006nasa}. A fixed questionnaire helps to identify potential overloads in a system. Several studies have used this questionnaire to evaluate the usability of VR software in various domains~\cite{feick2020virtual,zheng2012workload}.

\subsection{Usability Study of CIA Systems}
As CIA systems gain attention, there is a growing need to understand and address their usability challenges~\cite{kraus2021immersive}. Several critical factors, such as user interface design, interaction techniques, and collaboration mechanisms, can significantly impact the usability and effectiveness of these systems~\cite{ens2021grand}. Existing works offer crucial insights into IA systems, emphasizing opportunities for improvement in graphical perception, and the broad applicability in mixed reality~\cite{hoppenstedt2019applicability,whitlock2020graphical}.


A recent study has explored the challenges and opportunities within the CIA, particularly through the use of hybrid user interfaces~\cite{zagermann2023challenges}. 
Another study has addressed the importance of group awareness in collaborative environments that integrate VR and desktop platforms~\cite{seraji2022xvcollab}.

Despite these studies, there remains a gap in fully understanding how domain experts utilize CIA in real-world scenarios. Most of the existing studies focus on generic tasks or user groups causing a lack of detailed insights into specific domain challenges and workflow integration. This study aims to bridge this gap by exploring the unique usability challenges and opportunities of a CIA system, mostly used by domain experts. 
\section{Methodology}

This section outlines the comprehensive methodology used for the study. Experiment setup, participant demographics, experiment design, and evaluation metrics are briefly discussed to ensure clarity and reproducibility of the study.

\begin{table}[t]
\centering
\caption{PC configurations for the experiment settings.}\label{experiment_pc}
\begin{tabular}{l|l|l}
\hline
 &  PC 1  & PC 2\\
\hline
CPU &  Intel i7-13700K @3.4 GHz & Intel i9-9900X @3.4 GHz \\
RAM &  32GB at 4800 MHz & 32GB at 4800 MHz\\
GPU &  NVIDIA GeForce RTX 4070 Ti & NVIDIA GeForce RTX 3060 Ti\\
\hline
\end{tabular}
\end{table}

\subsection{Experiment Setup}
The experiments are conducted using ParaView Version 5.12.0 which is the latest public release during the study. The immersive visualization is facilitated through Meta Quest 2 headsets~\cite{quest2}. 

 
The hardware setup includes two high-performance PCs with the specifications mentioned in Table \ref{experiment_pc}. Both Meta Quest 2 headsets are attached to PCs using USB-C cables during the experiments. Remote participants are allowed to use their desktops equipped with Oculus Rift headsets in the experiments.

\subsection{Participants}
The study involves ten participants, which comprises a diverse group of individuals concerning age, gender, and educational background. The gender distribution includes eight men and two women. The age range is broad, with six participants between 20 and 29 years, one from 30 to 39, one from 40 to 49, and two aged between 50 and 59. 
Considering the highest education level, participants are also varied, with five holding bachelor's degrees, three with master's degrees, and two with doctoral degrees. It ensures a wide range of educational backgrounds from basic university education to advanced research qualifications. 
Having four participants with a history of motion sickness provides an opportunity to explore motion sickness-related issues in our experiments.

The group offers a rich blend of expertise in data visualization, including seven researchers, one developer, and two analysts. It helps to get a comprehensive understanding of the usability challenges from different professional viewpoints. Most of the participants are familiar with ParaView (7 out of 10), complemented by their proficiency with other data visualization tools, such as Python Matplotlib~\cite{matplotlib} and Seaborn~\cite{seaborn}, Tableau~\cite{tableau}, R~\cite{R}, MATLAB~\cite{matlab}, etc.


The participants' experience with VR technologies varies, with five classifying themselves as novices, three as intermediates, and two as experts. 
Participants specialize in data visualization across various fields such as scientific data, high-performance computing data, solar physics data, etc. However, only three participants have prior experience working on CIA projects.

\subsection{Experiment Design}
The study is structured into two distinct experiments. Each experiment is designed to evaluate different aspects of interaction within IA environments. Experiment 1 is conducted individually. Experiment 2 is structured to investigate collaborative dynamics by pairing participants to work together.

\begin{figure}[t]
\includegraphics[width=\textwidth]{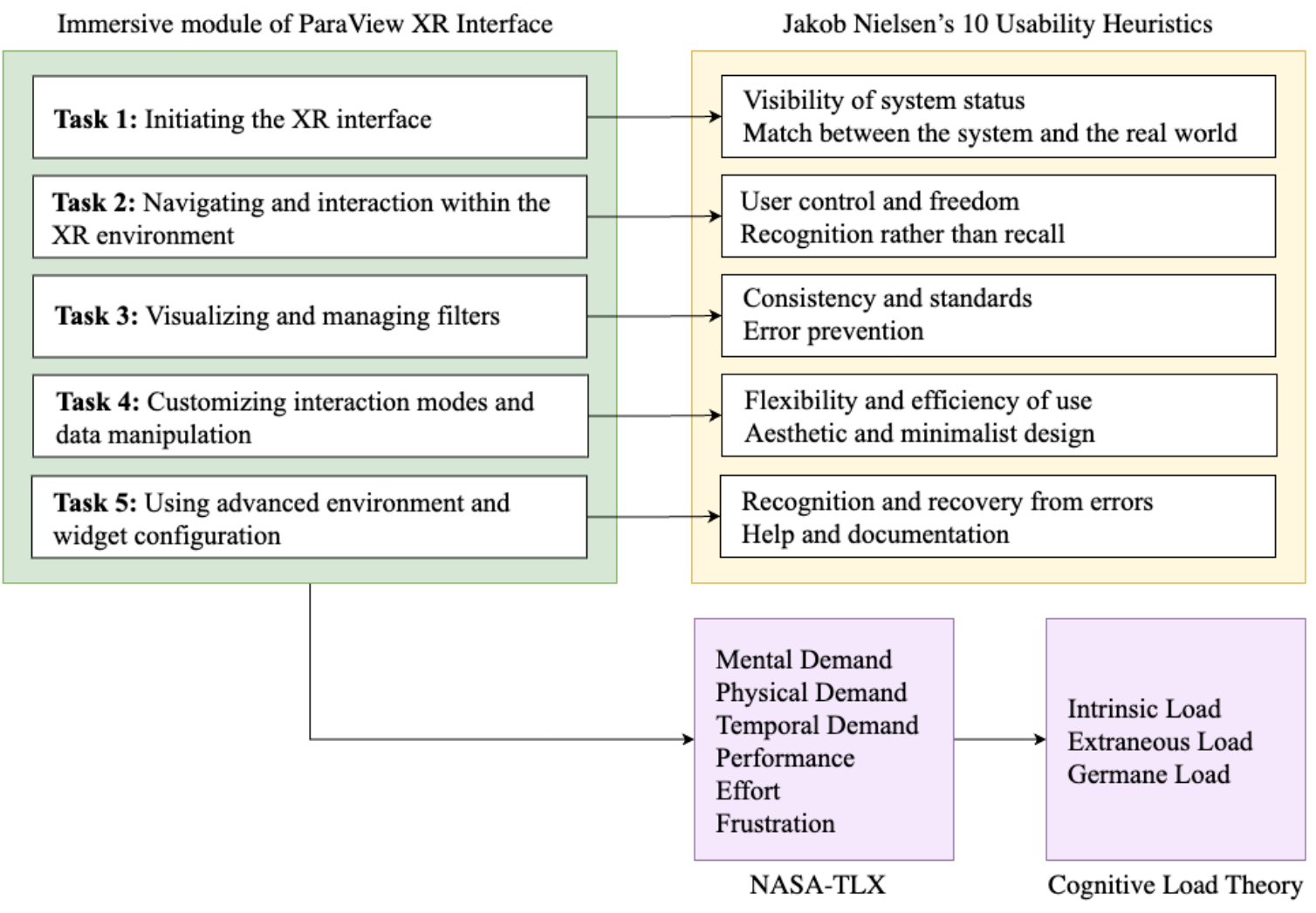}
\caption{Experiment design of the individual immersive analytics session.} 
\label{exp1}
\end{figure}

\vspace{0.05in}
\noindent{\bf Design of Experiment 1.} The first experiment, as illustrated in Figure~\ref{exp1}, is divided into five major tasks, each consisting of several subtasks that participants are required to demonstrate under the guidance of a research coordinator. Each task is designed to align with two of Jakob Nielsen's usability heuristics. Furthermore, the overall experience of the experiment incorporates CLT, with the application of the NASA TLX questionnaire to evaluate the cognitive demands imposed on participants. 

Task 1 focuses on initiating the XR interface, where participants are required to locate and activate the plugin, initiate the immersive experience from the ParaView desktop application, and familiarize themselves with the XR environment. This sequence is designed to ensure participants can transition smoothly from the desktop application to the XR environment. The task leverages the heuristic principle of \textit{``Visibility of System Status''} by ensuring users are aware of the system's status during the transition. 
Participants are allowed sufficient time to get familiarized with the virtual environment to examine the \textit{``Match between the System and the Real World''} heuristic. 

Task 2 involves participants using hand controllers to interact within the XR space. The subtasks include user control, icon identification, and menu navigation through different activities such as moving toward data sets and accessing menus. This task examines the heuristic of \textit{``User Control and Freedom''} allowing participants with the autonomy to navigate within the XR environment at their discretion. The task supports the heuristic of \textit{``Recognition Rather than Recall''} by enabling users to independently perform tasks after their initial exposure in the user menu.

Task 3 asks participants to use existing filters within the XR interface to visualize and interact with data. This task includes adjusting camera positions, interacting with data, enabling and disabling filters, and managing environmental settings, such as floor visibility. The design of this task adheres to the heuristic of \textit{``Consistency and Standards''} by maintaining familiar and consistent user interface conventions, which reduces the learning curve for the users. The task incorporates the heuristic of \textit{``Error Prevention''} to minimize potential user errors by expecting clear warnings before potential mistakes in the user journey.

Task 4 allows participants to engage directly with datasets through various interaction methods such as grabbing, interactive cropping, scaling datasets, and altering data coordinates. 
This task promotes the heuristic of \textit{``Flexibility and Efficiency of Use''} by assessing adaptability and personalization in the interaction modes that enhance user productivity. In addition, the task adheres to the heuristic of \textit{``Aesthetic and Minimalist Design''} by examining participants' views on distraction on the user interface. 

Task 5 involves participants in configuring the XR environment through various widgets, such as measuring the distance with scale and locating a certain data point in the virtual environment with a navigation panel.
The design of this task is integrated with the heuristic of \textit{``Recognition and Recovery from Errors''}
which provides users with the ability to identify and correct errors easily. 
Lastly, the task incorporates the heuristic of \textit{``Help and Documentation''}
by requiring users to locate and access assistance resources within the system.

These five tasks collectively aim to explore usability within the XR environment through the lens of CLT, using the NASA-TLX questionnaire as an evaluation tool. The questionnaire helps quantify the cognitive load imposed on participants by assessing mental demands, effort, and stress levels in different tasks. 
The insights gained from this questionnaire are critical to ensure an efficient and satisfying user experience in IA environments.

\begin{figure}[t]
\centering
\includegraphics[scale=0.48]{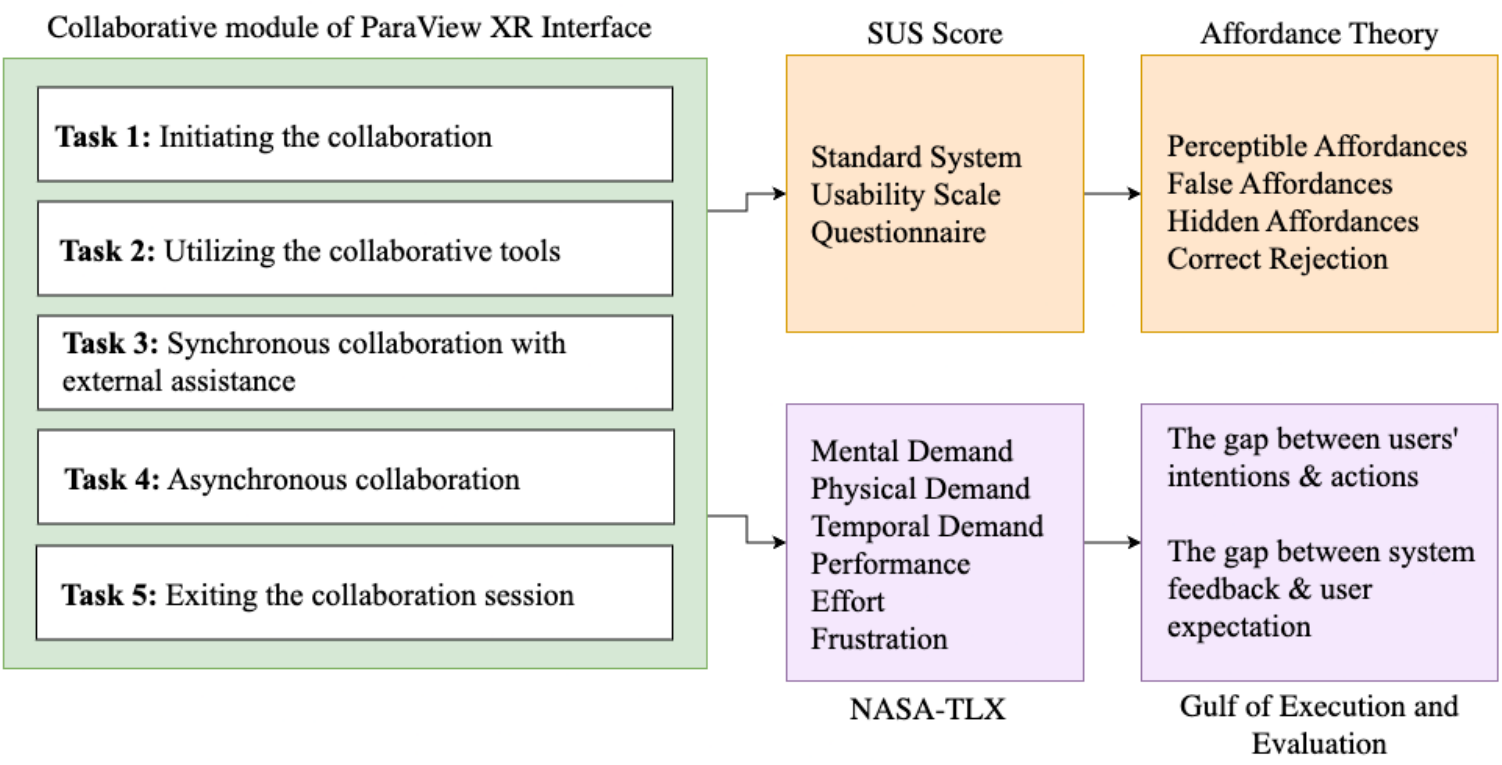}
\caption{Experiment design of the collaborative immersive analytics session.} 
\label{exp2}
\end{figure}

\vspace{0.05in}
\noindent{\bf Design of Experiment 2.} The second experiment, as illustrated in Figure \ref{exp2} consists of five major tasks, each comprising several subtasks that participants are instructed to demonstrate under the guidance of a research coordinator. 
This experimental setup utilizes affordance theory and the gulf of execution and evaluation to provide a thorough understanding of the user experience and efficiency of the CIA.

During task 1, participants are provided with a pvsm\footnote{The pvsm file format stands for ``ParaView State Machine'', saves the state of a ParaView session, including data sources, filters, views, and other settings.} file to launch their interaction with the collaborative module using ParaView's desktop application. It requires users to activate the plugin, followed by selecting the appropriate XR runtime. Participants are instructed to modify the default identifier to their name for better recognition within the collaborative environment. At the end, they connect to the collaborative server and launch the collaborative setting.

Task 2 requires participants to engage directly with the collaborative features of the system at the beginner level.  Users are asked to verbally recognize and call out the name or identifier of another collaborator within the virtual environment. They are instructed to make a gesture (waving) using the hand controller. Participants use the \textit{``Bring Collaborator Here''}
button to reposition the collaborator's avatar to a new location. Finally, they are asked to explore the availability of various collaborative tools, such as note-taking, data manipulation, and annotation.

Task 3 asks participants to attempt several synchronous collaborative tasks. Participants are tasked to perform various interactive operations such as grab, pick, and interactive crop on the data and show the modifications to their collaborator. Participants are also instructed to specifically point out areas of interest within the dataset to their collaborator, facilitating focused discussion. They are asked to use any existing widget (for example, ruler, navigation panel, etc.) of their choice to show its implementation to the collaborator. In the end, participants try to be involved in real-time communication between collaborators with and without external assistance.

Task 4 instructs participants to perform non-simultaneous collaborative work within the XR environment. Participants are asked to perform data manipulations, such as altering parameters or applying filters, and then save these adjusted states for future collaborators. The annotations and saved states are expected to communicate their analytical process asynchronously.

Task 5 of the experiment is straightforward. Participants are instructed to conclude the collaborative session, quit the XR environment, and disconnect from the collaborative server using the desktop application of ParaView.

\subsection{Experiment Procedures}
Participants are guided by research coordinators to perform specified tasks within a controlled environment during the experiments. Assistance is only provided if a participant fails a task three times, ensuring independent interaction with the system. After each experiment session, participants are asked verbally about any sensations of motion sickness. 

\vspace{0.05in}
\noindent {\bf Data Collection.}
User response is collected immediately following each experiment through a structured survey. The surveys are tailored to the specific requirements of the respective experiment session. 

Following the first experiment, the first survey consists of 16 questions. The initial 10 questions target specific usability heuristics defined by Jakob Nielsen. The remaining six questions are based on the NASA-TLX, using a scale from \textit{Low} (1) to \textit{High} (10).
This second part of the survey is designed to assess the cognitive load experienced by participants.

Following the second experiment, the second survey consists of 16 questions. The first 10 questions use the established SUS questionnaire. This standard questionnaire allows participants to rate their agreement on a scale ranging from \textit{"Strongly Disagree"} to \textit{"Strongly Agree"}, including options for \textit{“Disagree,”} \textit{“Neutral,”} and \textit{“Agree.”}. 
The following six questions, which are identical to the second part of Experiment 1, are based on the NASA-TLX focusing on evaluating the cognitive load concerning the gulf of execution and evaluation.

\section{Results and Discussions}
This section presents the results of two experiments aimed at assessing the usability of the ParaView XR interface. These experiments systematically evaluate the tool's performance and user interaction dynamics, with subsequent analysis based on established principles in HCI. 

\begin{figure}[t]
    \centering
    \includegraphics[scale=0.35]{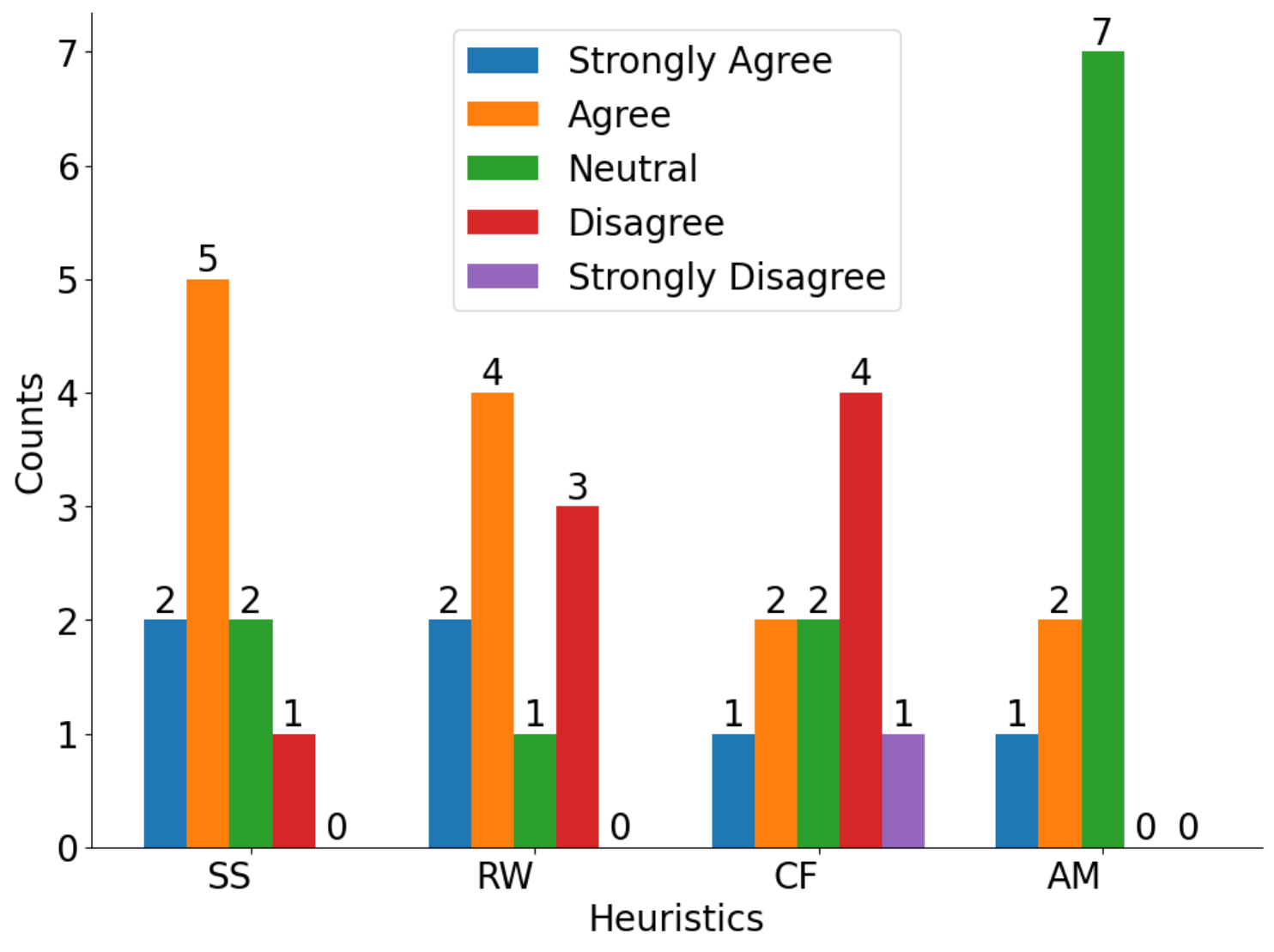}
    \caption{Results of experiment 1 based on Jakob Nielsen’s Heuristics. SS, RW, CF, and AM represent the visibility of system status, the match between the system and the real world, user control and freedom, and aesthetic and minimalist design, respectively.}
    \label{JN1}
\end{figure}

\subsection{Results of Experiment 1}
\begin{figure}[t]
    \centering
    \includegraphics[scale=0.3]{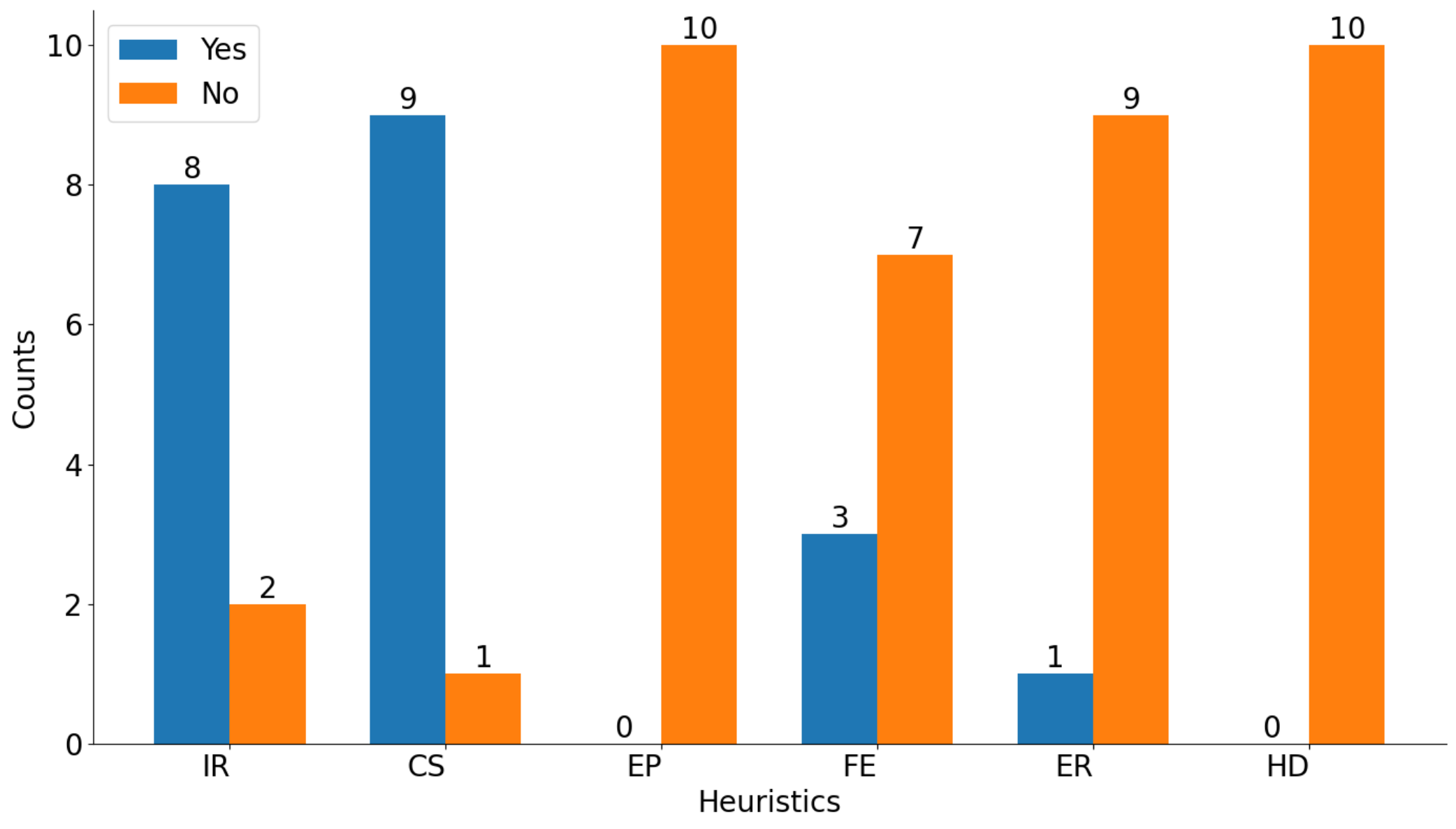}
    \caption{Results of experiment 1 based on Jakob Nielsen's Heuristics. IR, CS, EP, FE, ER, and HD represent recognition rather than recall, consistency and standards, error prevention, flexibility and efficiency of use, recognition and recovery from error, and help and documentation, respectively.}
    \label{JN2}
\end{figure}

\begin{figure}[t]
\centering
\includegraphics[scale=0.3]{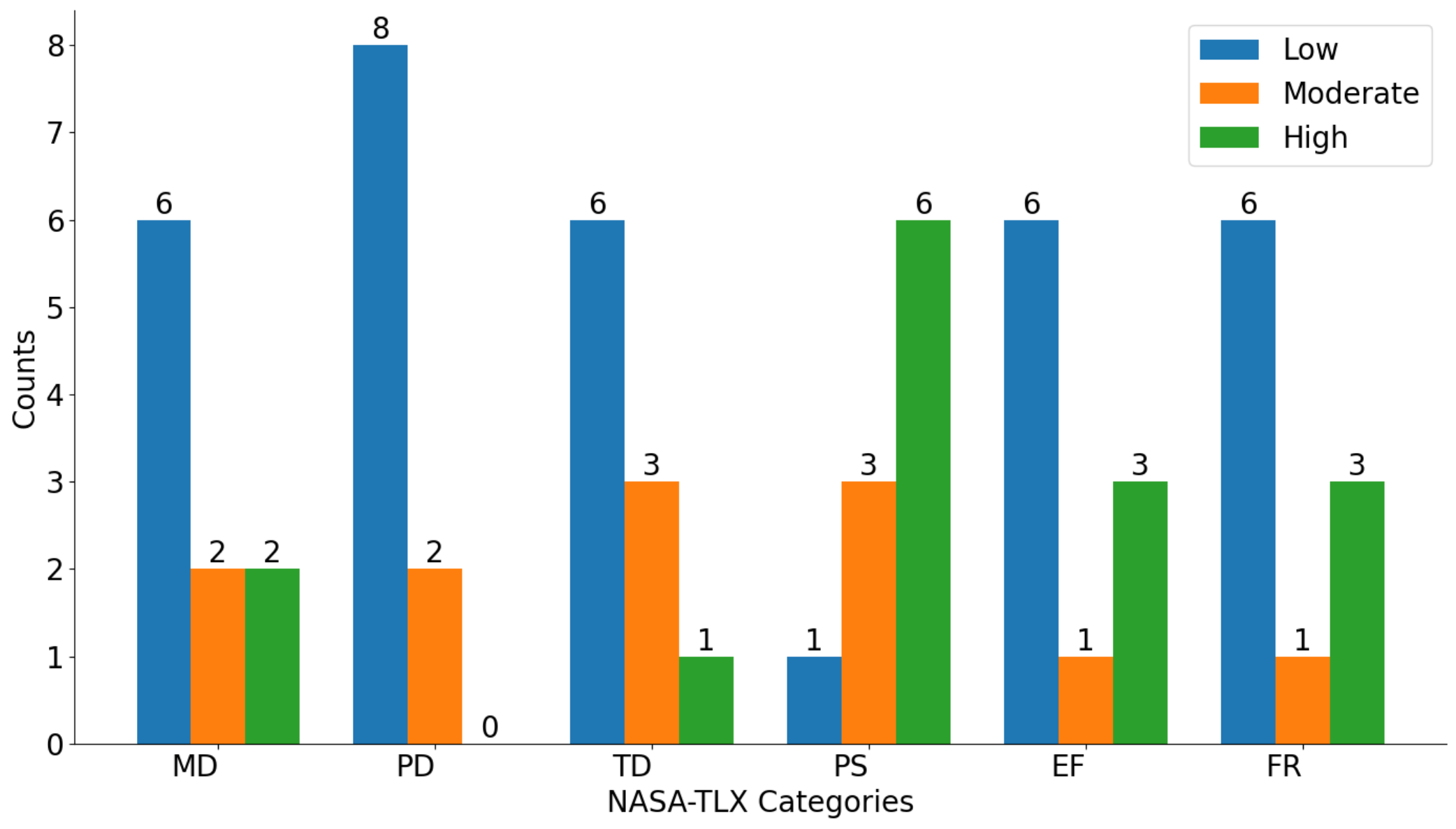}
\caption{Results of experiment 1 based on NASA-TLX. MD, PD, TD, PS, EF, and FR represent mental demand, physical demand, temporal demand, performance satisfaction, effort, and frustration, respectively.} 
\label{CL1}
\end{figure}

In the analysis of experiment 1, user responses are evaluated using Jakob Nielsen’s heuristics, as shown in Figure \ref{JN1} and \ref{JN2}. The cognitive load during the individual interaction is assessed through the NASA-TLX questionnaire, as shown in Figure \ref{CL1}.


\vspace{0.05in}
\noindent{\bf Visibility of System Status.} Most of the participants perceive the system to maintain the visibility of its status effectively, as denoted by \textbf{``SS''} in Figure \ref{JN1}. The system is successful in informing users about its status. Most users receive clear indicators during their interactions, successfully informing them of the transition from the desktop application to the XR interface.

\vspace{0.05in}
\noindent{\bf Match between the System and the Real World.} While the system partially meets real-world expectations, there are notable variances that impact user experience, as reflected by \textbf{``RW''} in Figure \ref{JN1}. The fact that 6 out of 10 respondents agree or strongly agree with the statement indicates that the system generally uses terms that are familiar to users. However, the users who disagree highlight a critical area where the system's terminology may not align well with user expectations. 

\vspace{0.05in}
\noindent{\bf User Control and Freedom.} It is essential to improve the undo and redo functionality to offer users more flexibility and control. Participants experience several challenges performing undo actions or reverting to a previous state, as presented by \textbf{``CF''} in Figure \ref{JN1}, drastically impacting user control in the virtual environment. This result indicates a notable deficiency in the system's response to user errors. 

\vspace{0.05in}
\noindent{\bf Recognition rather than Recall.} Identifying the icons and their functionalities within the ParaView XR interface indicates a strong alignment with the heuristic. The majority of participants, 8 out of 10, can identify the icons and their functionalities within the ParaView XR interface, as denoted by \textbf{``IR''} in Figure \ref{JN2}. The interface elements are recognizable without needing to recall information from memory, which supports an intuitive user experience. 

\vspace{0.05in}
\noindent{\bf Consistency and Standards.}
The majority of participants, with 9 out of 10 responses, find the interface actions and terms to be consistent throughout their use, as shown by \textbf{``CS''} in Figure \ref{JN2}. Consistency is crucial to improve user familiarity and reduce the learning curve. The consistent interface of ParaView XR interface allows users to rely on past experiences rather than relearning new interactions. 

\vspace{0.05in}
\noindent{\bf Error Prevention.} All 10 participants reported the absence of warnings or indicators before making mistakes, as reflected by \textbf{``EP''} in Figure \ref{JN2}. It is crucial to recognize errors by offering users clear warnings or indicators, assisting in the avoidance of potential issues. The lack of such features in the system suggests critical oversight in interface design, which can lead to increased user frustration. 

\vspace{0.05in}
\noindent{\bf Flexibility and Efficiency of Use.} The majority, consisting of 7 users share an opinion of a potential shortfall in the system's design to provide efficient tools that cater to diverse user requirements, as shown by \textbf{``FE''} in Figure \ref{JN2}. The goal of the IA environment is to streamline and enhance user interactions. A lack of an effective workflow acceleration mechanism can lead to increased operational times and reduced overall productivity. 


\vspace{0.05in}
\noindent{\bf Aesthetic and minimalist design.} The result highlights a predominantly neutral perspective, with 7 out of 10 users not particularly in favour or against the minimalism of the design, as referred by \textbf{``AM''} in Figure \ref{JN1}. Immersive applications are expected to provide an intuitive experience by focusing on essential elements and minimizing unnecessary information. The outcome supports that the user interface in the existing system adequately delivers information.

\vspace{0.05in}
\noindent{\bf Recognition and Recovery from Error.} A crucial aspect of maintaining user confidence and minimizing frustration during interaction is helping users recognize and recover from errors efficiently. However, 9 out of 10 participants indicate that the system does not help them in this regard, as denoted by \textbf{``ER''} in Figure \ref{JN2}. The system lacks the necessary feedback mechanisms or instructional guidance to alert users about errors and guide them towards solutions effectively. 

\vspace{0.05in}
\noindent{\bf Help and Documentation.} A  system should offer accessible and useful documentation or on-the-spot help to assist users in resolving issues or uncertainties. However, the result shows unanimous user feedback, indicating that no help or documentation is provided when stuck on a task in the system, as reflected by \textbf{``HD''} in Figure \ref{JN2}. The absence of such support in the system can lead to increased frustration, as users are left to troubleshoot issues without guidance.

\vspace{0.05in}
\noindent{\bf Analyzing the Cognitive Load.} The responses from the NASA-TLX questionnaire, as shown in Figure \ref{CL1}, indicate a lower trend of mental, physical, and temporal demands experienced by participants during the individual task. It suggests that the tasks do not require significant cognitive resources and physical effort. A significant majority feel satisfaction with their performance while experiencing the immersive experience individually. Although the task is challenging for some, the majority of participants report for low level of effort and frustration. Overall, participants face fewer challenges and cognitive load while experiencing the ParaView XR interface individually without any requirement of collaboration.

\subsection{Results of Experiment 2}
In the analysis of results from experiment 2, user responses will be evaluated by combining the SUS and the affordance theory. In addition, the execution and evaluation will be assessed using the NASA-TLX questionnaire, providing information on user expectations in the collaborative environment.

\begin{figure}[t]
\centering
\includegraphics[width=\textwidth]{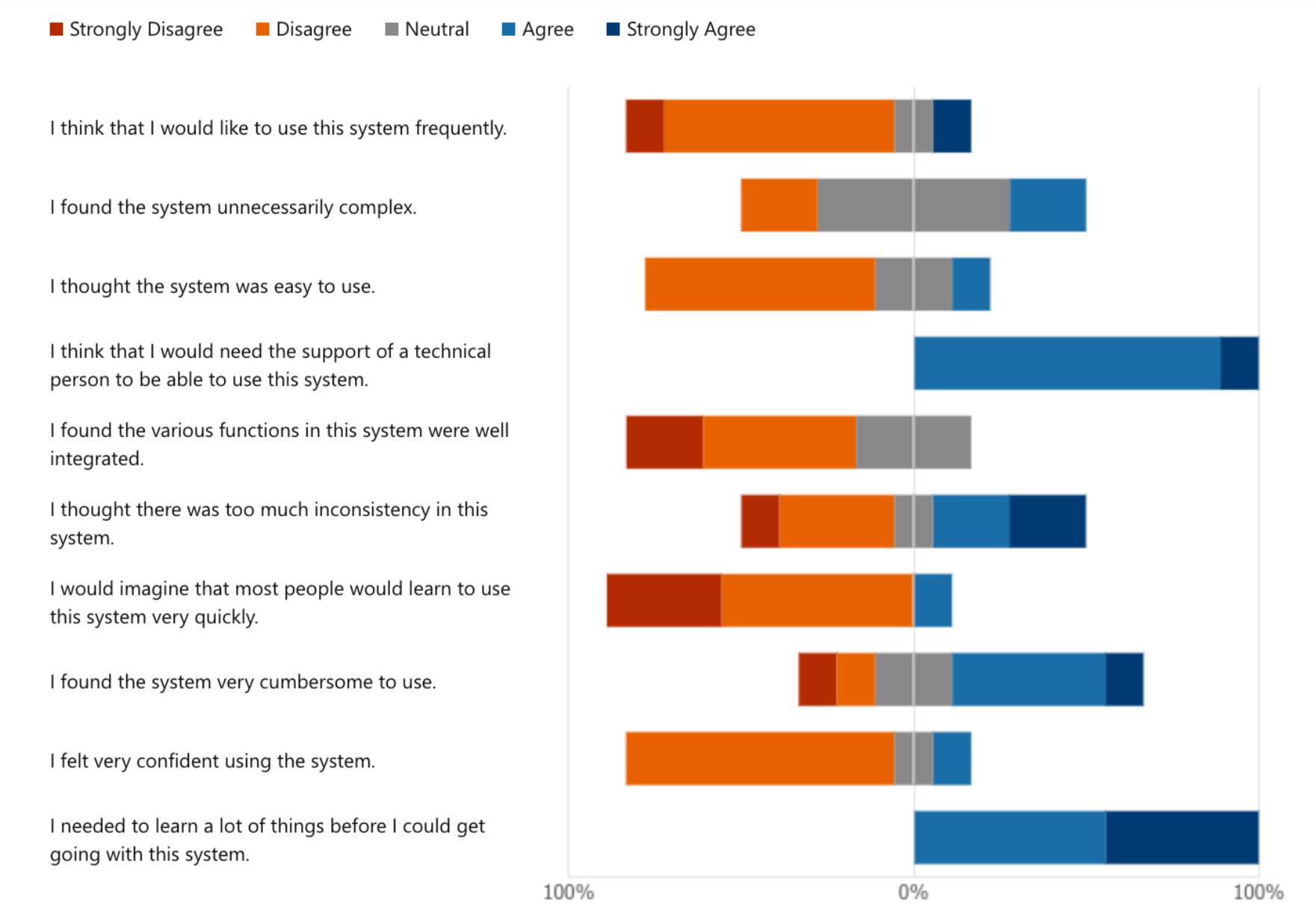}
\caption{Results on experiment 2 based on the SUS questionnaire.} 
\label{exp2_sus}
\end{figure}
\vspace{0.05in}
\noindent{\bf System Usability Scale} is used to assess the usability of the CIA system used in this experiment, as shown in Figure \ref{exp2_sus}.
The SUS score obtained for the collaborative module of the ParaView XR interface is \textbf{32.78}, which significantly indicates the challenges encountered by participants during collaborative sessions. This SUS score marks below the threshold of 68 which is considered as the benchmark for above-average usability. This low score reflects substantial usability barriers that hinder effective user interaction and collaboration.

\vspace{0.05in}
\noindent{\bf Task-wise Analysis based on Affordance Theory.} The underlying cause of the low score in SUS is evident in the achievement of the tasks in experiment 2. 

In task 1, participants effectively engage with the collaborative module by activating the plugin, selecting the appropriate XR runtime and customizing the default identifier, demonstrating an example of perceptible affordance. However, they encounter difficulties in launching the collaborative settings due to an inactive button, indicating a false affordance in the system. 

In task 2, participants experience perceptible affordance by successfully recognizing other collaborators and performing gestures using hand controllers. However, while using the \textit{``Bring Collaborators Here''} button to reposition avatars, participants face false affordance in the system causing unexpected overlap of the avatars. The system’s inability to accurately locate users in both real-world and virtual environments is observed during the experiment, highlighting an inadequacy in environmental knowledge and contextual understanding. Participants faced a limitation of collaborative tools for data manipulation or annotation, which is a design gap in the system. 

In task 3, though participants perform the interactive operation, they are unable to share it with their collaborators because of the asynchronous nature of the system. All participants fail to point out areas of interest within the data set and showcase the implementation of existing widgets to their collaborators due to the same challenge of inability to synchronize. The absence of any communication method in the system, such as voice or text, hinders the ability to synchronize. All of these issues are identified as a design deficiency in the capabilities of the system.
   
In task 4, participants encounter false affordance in the system while engaging in nonsimultaneous collaborative work in the CIA system. Despite completing the required subtasks, participants fail to save the modified states to be shared with future collaborators due to system constraints. The lack of role-based access control in the system prevents effective management of user roles, such as collaborator, reviewer and observer, causing challenges to ensuring seamless collaboration among participants.

In task 5, participants face challenges when trying to disconnect from the collaborative server within the virtual space. There is no visible button on the XR interface to disconnect directly from the collaborative server, reflecting a hidden affordance in the system.

In general, the system contains numerous false affordances, hidden affordances, and design deficiencies which result in significant usability challenges.


\begin{figure}[t]
\centering
\includegraphics[scale=0.3]{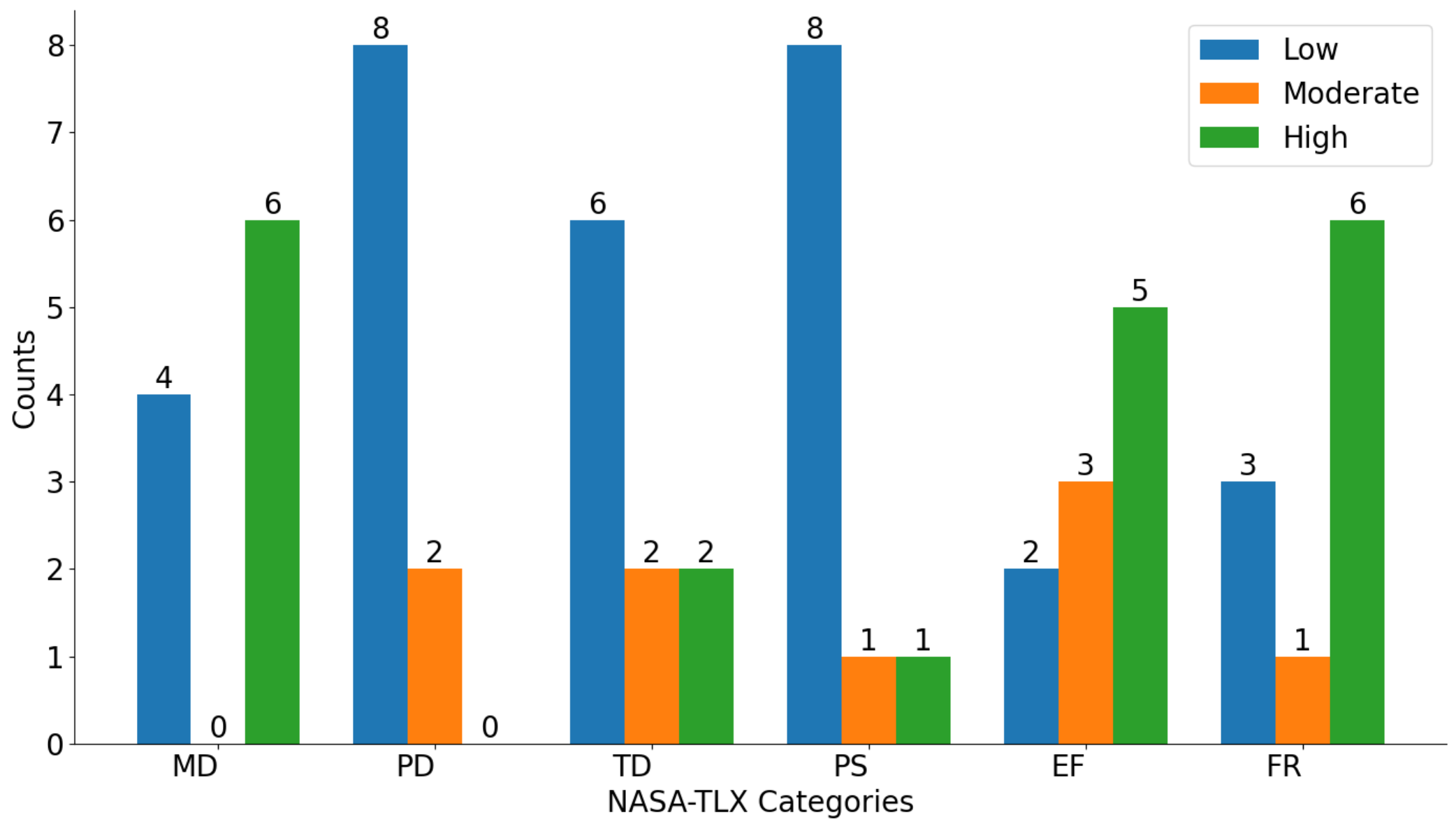}
\caption{Results of experiment 2 based on NASA-TLX. MD, PD, TD, PS, EF, and FR represent mental demand, physical demand, temporal demand, performance satisfaction, effort, and frustration, respectively.} 
\label{Exp2_Nasa}
\end{figure}
\vspace{0.05in}
\noindent{\bf Analyzing Gulf of Execution and Evaluation.} The responses from the NASA-TLX questionnaire, as shown in \ref{Exp2_Nasa}, indicate varying levels of mental, physical, and temporal demands experienced by participants. Mental demand appears to be high for the majority of the participants suggesting tasks require a significant level of cognitive processing. Temporal demand varies from user to user due to inconsistent performance and unexpected behaviour of the system. There is a negative trend of performance satisfaction, effort, and frustration among participants in the collaborative experience compared to the individual experience.

The result highlights a significant gulf in execution, where the actions required by the system do not align with the mental models of the participants. This suggests that efficiently translating intentions into actions is a challenge in the CIA system. The recurring system lags and unexpected behaviour worsen the gulf of evaluation, as these issues hinder users’ ability to interpret system responses accurately. The observed decline in performance satisfaction highlights disparities between expected and actual system performance. The increased frustration and required efforts, in collaborative settings, compared to individual scenarios, reflect the gulfs of execution and evaluation becoming more noticeable in the CIA system.



\subsection{Key Findings in the CIA Module}
This section presents the key findings derived from the experimental evaluation of the CIA module of the ParaView XR interface.
\begin{itemize}
    \item The CIA module lacks support for synchronous collaboration and communication techniques for users.
    \item The CIA system cannot assign user roles among collaborators, affecting uninterrupted user experience.
    \item The CIA system is ineffective in resolving conflict resolution in shared spaces between collaborators due to insufficient contextual awareness. 
\end{itemize}

\section{Recommendations}
This section shares a set of strategic recommendations aimed at enhancing the capabilities of the ParaView XR interface. These recommendations are specifically tailored for domain experts engaged in complex data visualization tasks.

\vspace{0.05in}
\noindent{\bf Real-time Synchronization in Collocated and Distributed Environments.} Real-time synchronization is important in CIA. It becomes even more crucial especially when domain experts work in both collocated and distributed settings~\cite{du2018zero,radu2021survey}. The goal is to ensure that all participants have a consistent and up-to-date view of the data, irrespective of their physical location, facing the challenge of minimizing latency across varied network conditions~\cite{fraser2000revealing}.

One approach to address this issue is implementing a differential synchronization algorithm~\cite{fraser2009differential} which allows the system to transmit the changes made in the shared environment instead of the entire data set. 
Adding a local caching scheme can significantly enhance real-time synchronization in collocated and distributed environments by reducing the frequency of data transmissions~\cite{zhang2022sear}.


Adaptive synchronization~\cite{ginosar1998adaptive} strategies based on the type of data and the specific collaboration context can further optimize the synchronization process. For example, prioritizing the synchronization of critical data or using delta encoding techniques to minimize the amount of transmitted data can be utilized~\cite{tan2020exploring}.

Furthermore, modern network protocols such as WebRTC~\cite{rfc7478} can enhance the communication infrastructure of the CIA, maintaining a secure and reliable channel to transmit data~\cite{petrangeli2018improving}. 
It can be an ideal choice for handling the complexities of interactive data visualizations involving high-dimensional data sets because it supports real-time media and data channels~\cite{nakazato2024webrtc}.

\vspace{0.05in}
\noindent{\bf Dynamic Role-based Access Control (RBAC) in CIA.} Implementation of RBAC~\cite{sandhu1998role} can improve security and ensure appropriate access permissions in CIA environments. Traditionally, RBAC models statically assign permissions based on predefined policies~\cite{ferraiolo2001proposed}. However, permissions may need to be adjusted in real-time for a dynamic collaborative environment based on various contextual factors, such as user roles, location, current activities, etc.~\cite{park2001role}. 


Another advanced approach can be attribute-based access control~\cite{hu2015attribute} which allows access based on attributes associated with users, resources, and environmental conditions. It provides more flexible access control policies which are dynamically evaluated and enforced based on the changing attributes within the collaborative environment~\cite{servos2017current}. Integrating advanced machine learning algorithms can facilitate the automatic adjustment of access controls based on behaviour patterns~\cite{afshar2021incorporating}.


\vspace{0.05in}
\noindent{\bf Integrating Environmental Context Awareness into Simultaneous Localization and Mapping (SLAM).}
To enable the construction of 3D maps and the tracking of user positions and orientations within the environment for CIA applications, SLAM~\cite{durrant2006simultaneous} algorithms are essential~\cite{covolan2020mapping}. Sole dependency on sensor data, such as depth cameras makes traditional SLAM algorithms ineligible for the collaborative environment. This limitation can be addressed by integrating environmental context awareness into the mixed reality-based systems~\cite{dasgupta2020towards}.


By leveraging computer vision and machine learning techniques, semantic SLAM~\cite{zhang2018semantic} algorithms can identify and recognize specific objects, surfaces, or environmental features. It uses this semantic information to improve the accuracy of mapping and localization. 
Deep learning-based SLAM~\cite{su2022real} approaches leverage deep neural networks to improve the reliability of SLAM algorithms. It can perform better in challenging environments or when dealing with dynamic scenes. Utilizing recurrent neural networks (RNNs) or long short-term memory (LSTM) networks ensures accurate tracking and mapping in dynamic scenes~\cite{alam2023review,azzam2020stacked}. Convolutional neural networks (CNNs) can assist in feature extraction and improve the correctness of feature tracking~\cite{tateno2017cnn}. 

\section{Conclusion}
This study conducts an extensive experiment on ParaView's capabilities in facilitating CIA. It uncovers important insights into both the platform's strengths and the areas needing improvements. Our findings from this research demonstrate that the collaboration mode of ParaView currently presents significant usability issues that hinder effective teamwork. Considering the unsatisfactory sentiment among users, the CIA system demands an urgent need for enhancements. As CIA becomes increasingly ubiquitous in professional settings, the simplicity of tool navigation and the clarity of the feedback provided is paramount for user experience.

Usability in the CIA is not a feature anymore. When the usability barriers are high, it can obstruct user engagement, leading to underutilization of potential technology. Therefore, it is essential to fully leverage the collaborative capabilities and innovative potential of immersive analytics tools by optimizing their usability. Developers, practitioners and the research community can significantly enhance the usability and effectiveness of the CIA by considering the recommendations. We wish our research paves the way for more insightful, inclusive, and impactful CIA endeavours.


\bibliographystyle{splncs04}
\bibliography{bibliography}

\appendix
\section{Ethics}
This study has been approved by the Institutional Review Board (IRB) and does not raise ethical issues.

\end{document}